\documentclass{elsart}
\usepackage{graphicx}
\usepackage{amssymb,amsmath,array}

\def\exp{\mathop{\rm exp}\nolimits}
\def\w{\widetilde}

\begin{document}

\begin{frontmatter}
\title{Slow dynamics in a model of\\ the cellulose~network}
\author{O.V. Manyuhina\corauthref{cor}}
\corauth[cor]{Corresponding author. Tel.: +31\,24-365\,28\,49}
\ead{o.manyuhina@science.ru.nl}
\author{A. Fasolino}
\author{M.I. Katsnelson}
\address{Institute for Molecules and
Materials, Radboud University Nijmegen, Toernooiveld~1, 6525 ED
Nijmegen, The Netherlands}

\begin{abstract}
We present numerical simulations of a model of cellulose consisting of long stiff rods, representing cellulose microfibrils,  connected by stretchable crosslinks, representing xyloglucan molecules, hydrogen bonded to the microfibrils. Within a broad range of temperature the competing interactions in the resulting network  give rise to a slow glassy dynamics. In particular, the structural relaxation described by orientational correlation functions shows a logarithmic time dependence. The glassy dynamics is found to be due to the frustration introduced by the network of xyloglucan molecules. Weakening  of interactions between rod and xyloglucan molecules results in a more marked  reorientation of cellulose microfibrils, suggesting a possible mechanism to modify the dynamics of the plant cell wall.
\end{abstract}

\end{frontmatter}

\section{Introduction}

Cellulose is the most abundant organic material in nature
and it is the basic structural component of the plant cell
walls. In the higher plants, cellulose crystalline microfibrils,
5-15~nm wide and \mbox{1-5}~$\mu$m long, are embedded in a
matrix of hemicellulose,  pectin polysaccharides and  proteins~\cite{cosgrove:1997}. Bacterial cellulose composite, produced by {\it Acetobacter xylinum} growing in a medium containing xyloglucan (hemicellulose), is a pure cellulose--xyloglucan network with degree of crystallinity
around 75\%~\cite{sullivan:1997,iwata:1998}. Xyloglucan molecules are bound by hydrogen bonds to
the amorphous surface of different cellulose fibrils creating crosslinks
between them, which imparts elastic properties to the network. 

The dielectric properties of bacterial cellulose were studied in
the frequency range 100~Hz--1~MHz and at temperatures between
100 and 440~K~\cite{baranov:2003}. A broad
peak of dielectric absorption was found for $T<260$~K, indicating a
wide distribution of  relaxation times. Creep experiments on
cell walls of higher plants and on composite material from {\it
Acetobacter xylinum} have shown a logarithmic time dependence of
elongation~\cite{nieuwland:2005}. Moreover, the authors have shown that, 
in presence of the cell wall loosening lipid transfer proteins (LTP), the elongation remains logarithmic but with a higher creep rate. These slowing down of relaxation and logarithmic time evolution are dynamical features typical of glassy systems~\cite{binder:1986,gotze:1992,gotze:2002,mallamace:2000}. Evidence of glassy behaviour in biologically
relevant materials may be of general importance. Schr\"{o}dinger in
his seminal book~\cite{whatislife} described the building blocks
of life as ``aperiodic crystals'',  orderly structures withdrawn 
from the disorder of heat motion. Much later, the general concept of glassy state has been developed~\cite{binder:1986,mezard:1987,toulouse:1977}) as a state with infinitely many local
metastable configurations. Such a glassy state is prone to minor changes of environment since even small external actions can transfer the system as a whole into a different metastable state, while never reaching thermodynamic equilibrium. Interestingly, the transition to a glassy state found in saccharides at low temperature~\cite{nagasawa:2004} has been suggested  to be the 
key factor in the protection of biological tissue against freezing.
For these reasons, it is plausible to assume that living organisms should make  use
of  substances with a broad variety of similar structures depending on external conditions,
as offered by glassy materials. Evidence of slow glassy dynamics of the 
cytoskeleton of living cells was recently observed experimentally~\cite{fabry:2001,bursac:2005,deng:2006}. The
simplest and, at the same time, one of the most important organic
materials, cellulose, is a good starting point to investigate this possible connection theoretically.

The physical mechanism which is responsible for glass formation is
the presence of {\it frustration}, i.e. the incompatibility between
the locally preferred order and global constrains. In
other words, the energy of the system cannot be minimized 
by optimizing all local atomic interactions~\cite{toulouse:1977,tarjus:2005}.

In the cellulose--xyloglucan system, frustration gives rise to defects in
the microfibril organisation due to  long-range
xyloglucan--xyloglucan interactions. To simulate this
phenomenon, we propose a model that describes 
cellulose microfibrils and xyloglucan molecules at the mesoscopic, coarse grained, level. 
Correlation functions of the microfibril orientation calculated within this model display 
the slow logarithmic behaviour typical of glasses
over two time decades. Similar logarithmic decay of correlators  has been
observed in Monte Carlo simulations for
spin-glasses~\cite{binder:1986},  
colloids and micellar particles~\cite{mallamace:2000,chen:2003,puertas:2002}, and overcooled liquids~\cite{novikov:2003}. This slow dynamics is often explained 
in terms of mode-coupling theory~\cite{gotze:1992,gotze:2002}.

In the following we present the model and the results of
molecular dynamics (MD) simulations for the cellulose
network, analyze the structure and dynamics identifying the glassy
behaviour, and point out the crucial role of the xyloglucan
network for the slow dynamics of cellulose microfibrils.


\section{Model and Simulation Technique}
\begin{figure*}
\centering
\includegraphics[width=8.25cm]{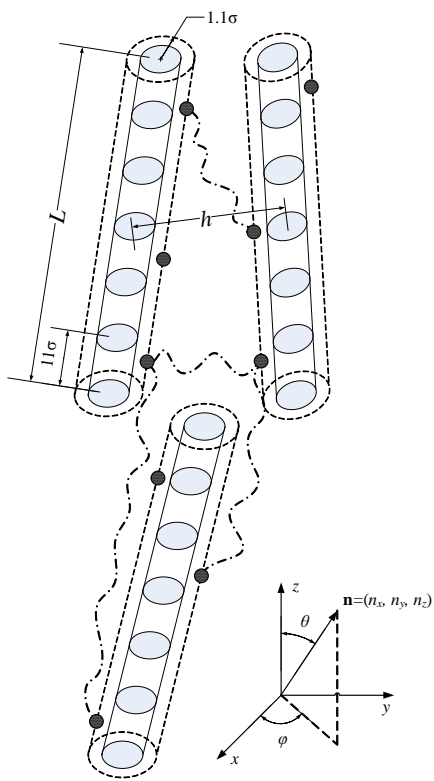}
\caption{Schematic representation of our model: each rod are formed by seven beads (grey spheres). Each x-particle (black spheres) interacts with the rod beads and with other two x-particles (see text) as indicated by the dashed-dotted wavy lines. The ratio between rod beads and x-particles is $7\text{:}3$.  The rods are closely packed, i.e. the equilibrium length of rod $L=67.3\sigma$ is significantly larger than the equilibrium space between them $h=4.5\sigma$. The rod director $\bf n$ is also indicated in cartesian and polar coordinates.}
\label{fig:model}
\end{figure*}

We propose the coarse grained atomistic model illustrated in Fig.~\ref{fig:model} 
to simulate bacterial cellulose~\cite{cosgrove:1997}, 
a prototype of cellulose in higher plants. The  model consists of long stiff rods, each constituted by seven beads, representing cellulose microfibrils,  
connected by a network of stretchable crosslinks, representing xyloglucan molecules.


In our coarse-grained model, there are four types of interactions
between: 1)~beads within a rod~(b-b), 2)~beads in different rods~(r-r), 
3)~xyloglucan particles and rod beads~(x-r)
and 4)~xyloglucan particles~(x-x).
We have chosen these interactions 
with the following criteria: 
i)~the x-r bonding sets the scale of energy to that typical of a hydrogen bond,
ii)~the rods have to be stiff, impenetrable, very long compared 
to typical bonding lenghts, and the interactions between them are dominated by the interactions via xyloglucan particles,
iii)~the xyloglucan has to form a  disordered  
network connecting different rods with the possibility of bond stretching, breaking 
and formation. This is less straightforward than the previous steps and will be described in some details later~on.

We describe most interactions by a Lennard-Jones potential of the form:
\begin{equation}\label{eq1.1}
U_{{\rm LJ},s}(r) = 4\varepsilon_s\bigg[\bigg(\frac{\sigma_s}
r\bigg)^{12}-\bigg(\frac {\sigma_s} r\bigg)^6\bigg]+C_s,
\end{equation}
where $s = \{\text{b-b, x-r, r-r, x-x}\}$ labels the different interactions. 
The minimum of the potential, -$\varepsilon_s$,  occurs at $r_{{\rm eq},s}= 2^{1/6}\sigma_s$.
The constant $C_s$ is chosen in such a
way that  the potential vanishes at $r =R_{c,s}$, where
$R_{c,s}$ is a cut-off radius and $U_{{\rm LJ},s}=0$ for $r > R_{c,s}$. The parameters are given in Table~\ref{tab1}
in units of the characteristic energy and length of x-r interactions, $\varepsilon\equiv\varepsilon_{\rm x\text{-}r}$ and 
$\sigma\equiv\sigma_{\rm x\text{-}r}$. 
For bacterial cellulose, these units can be assumed as $\varepsilon\approx 20$~kcal/mol $\approx 10^{4}$~K, which
describes a strong hydrogen x-r bonding~\cite{cosgrove:1994}, and $\sigma\approx 15$~ nm that gives the typical radius of the microfibrils.
In these units, the equilibrium length of each rod corresponds to the 
typical length of cellulose microfibrils, $L=1$~$\mu$m with an equilibrium distance between
them $h=67$~nm (Fig.~\ref{fig:model}). Moreover, the cut-off radius of $U_{\rm LJ,x\text{-}r}$ is chosen in such a way that each xyloglucan particle interacts with no more than one rod $R_{c,{\rm x\text{-}r}} = \frac12 r_{{\rm eq,r\text{-}r}}$ (Table~\ref{tab1}).

In order to keep the rods rigid, we add to $U_{\rm LJ,b\text{-}b}$ a simple bending potential
between the nearest beads within a rod given as
\begin{equation}\label{eq1.2}
U_{\rm B} = A(1+\cos\alpha),\quad \cos\alpha_i =\frac{({\bf
r}_{i-1}- {\bf r}_i,{\bf r}_{i+1}-{\bf r}_i )} {|{\bf r}_{i-1}-{\bf
r}_i|\cdot|{\bf r}_{i+1}-{\bf r}_i|},
\end{equation}
where $A \gg1$ (Table~\ref{tab1}), ${\bf r}_i$ is the position
of the bead $i$, with $i = 2\text{--}6$. This strong potential allows to decrease the cut-off radius $R_{c,{\rm b\text{-}b}}$ of $U_{\rm LJ,b\text{-}b}$ and lower the computational time. 


We come now to the interactions between xyloglucan particles, which play the role of crosslinks between rods. To obtain a connected network of rods with effective long-range interactions, we require each xyloglucan particle to interact with two others. These triplets of connected particles are chosen once and for all at the beginning of the simulation as described in the following.

\begin{figure*}
\centering
\includegraphics[width=8.25cm]{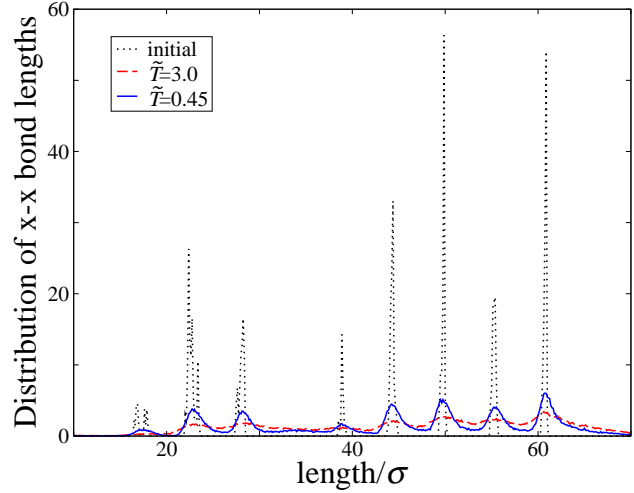}
\caption{Bond length distribution of connected x-x triplets.  Dotted line: initial sample  providing the $r_{\rm eq}$-distribution (see text); dashed and solid lines: equilibrated system around room temperature ($\w{T}=3.0$) and at low temperature ($\w{T}=0.45$), respectively.}
\label{fig:xyl_distr}
\end{figure*}

We place three x-particles per rod at the equilibrium distance of 
$U_{\rm LJ,x\text{-}r}$ from the second, fourth and sixth bead of each rod.  
For each x-particle we choose a first partner  as the farthest x-particle within a radius $L$ that is not yet assigned as partner of another particle.
We repeat this procedure to assign the second partner. 
Once all triplets have been assigned, the distances between them in the initial 
structure are defined as equilibrium distance $r_{\rm eq}$ for the additional interaction between
particles within a triplet described by a Morse potential
\begin{equation}\label{eq1.3}
U_{\rm M} = D\big(1-e^{-b(r-r_{\rm eq})}\big)^2,
\end{equation}
with parameters $D,b$ specified in Table~\ref{tab1}.
This means that every pair interacts through a potential with minimum at a different 
value of $r_{\rm eq}$. 
As shown  in Fig.~\ref{fig:xyl_distr}, the initial $r_{\rm eq}$-distribution contains seven sharp peaks. We further assume that a bond between x-x pairs can be broken if
$U_{\rm M} \geqslant 0.9D$, which  introduces a cut-off
radius $R_c^{\rm M}$ for x-x interactions
\begin{equation}\label{eq1.4}
R_c^{\rm M} = r_{\rm eq}-\frac{\ln (1-\sqrt{0.9})}{b}\simeq r_{\rm eq}+\frac{2.97}b.
\end{equation}
The resulting network is very robust and less than 6\% of the x-x bonds break at all studied temperatures. The long-range interactions between xyloglucan particles introduce frustration in the orientation of the rods, that in their absence, would energetically prefer to form a crystal of parallel rods. Although there is a large arbitrariness in our construction of the interactions, we can assume that frustration due to competing interactions is a quite general feature of non crystalline multicomponent systems, (like most living matter).

\begin{figure*}
\centering
\includegraphics[width=8.25cm]{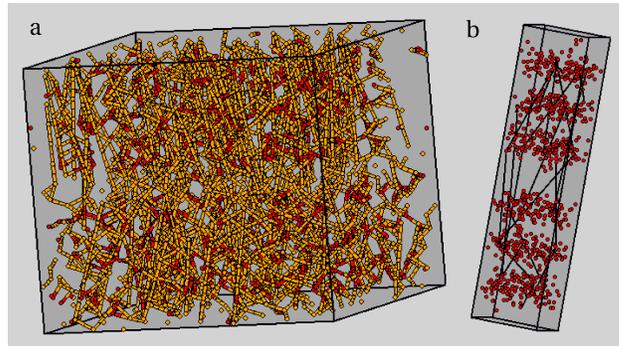}
\caption{Snapshot of the system at $\w{T}=0.45$ in two representations. (a) Yellow balls and sticks: rods; red balls: xyloglucan particles. 
One can distinguish two layers of rods roughly oriented along the $z$ direction. For illustration purposes the $x$ and $y$ sides of the box are multiplyed by five. (b)  The simulation box is shown with real relative dimensions. Only the x-particles (red balls) and a few of the bonds (black lines) within triplets are shown.}
\label{snapshot}
\end{figure*}

Simulations of glasses have to be conducted possibly over several
decades in time. However  this fact, together with the complexity of the
interatomic potentials, limits the size of the system. In our
simulation we use ${N_{\rm rod} = 288}$ rods, located in parallelepiped with 6 unit cells in the $x$ and $y$ directions and 14 in the $z$ direction. In each unit cell, defined by vectors
$$
{\bf i}=(6\sigma,0,0),\quad {\bf j}=(0,6\sigma,0),\quad {\bf k}=(0,0,11\sigma),
$$
we put four beads at $(0.25,0.25,0.25)$, $(0.75,0.75,0.25)$, $(0.25,0.75,0.75)$, $(0.75,0.25,0.75)$. We form the rods by letting the beads interact with each other in groups of seven along the $z$ direction, which implies two layers of densely packed rods along the $z$ direction. 
In Fig.~\ref{snapshot}a we show a typical snapshot of the equilibrated system where the two layers of rods and the x-r bonds are visible, whereas in Figure~\ref{snapshot}b we show only the bonds between a few xyloglucan triplets.

In essence, our system is like a liquid crystal with the unit vector of rod orientation ${\bf n}$ (Fig.~\ref{fig:model}), playing the role of order parameter. 
The constructed configuration, composed of $N = 2880$ particles, was 
equilibrated at high temperature $T=0.16\varepsilon/k_B$ in the microcanonical
$NVE$ ensemble.
We integrated the equations of motion rescaled by the mass of the
rod bead taken to be  $m=10^{-22}$~kg, which is twice the  mass of xyloglucan particles, by
means of the velocity Verlet algorithm. The time step in units of
$\tau=\sqrt{m\sigma^2/\varepsilon}\approx 0.5$~ns was chosen $\Delta t=0.032$,
the shortest period of vibrations between x-r particles being $30\Delta t$.
Then, the system was quenched with rate $3\cdot 10^9$ K/s and studied in the $NVT$ ensemble, and studied at $T = (3.0, 1.9, 0.9, 0.45, 0.23, 0.11)\cdot10^{-2}\varepsilon/k_B$, using the Nose--Hoover chain thermostat~\cite{FrenkelSmit}. The masses of the thermostat chains were defined by
$Q_1 = NT/\omega^2$ and $Q_2=T/\omega^2$ with $\omega=6.7$, the
characteristic frequency of x-r vibrations. For convenience, in the following, we give
the values of temperature  in reduced units as $\w{T}= 10^2 k_BT/\varepsilon$. Note that, $\w{T}=3.0$ corresponds to room temperature.

In the following sections, we study separately the effect of
temperature and the role of intermolecular interactions by
comparing systems with weak x-r bonds ($\tilde{\varepsilon}_{\rm x\text{-}r}=0.1\varepsilon$) and
broken x-x bonds ($D=0$) to the original one (Table~\ref{tab1}).


\section{Static fingerprints of the glassy state}


We describe the static structure by calculating the distribution function
of the polar angle~$\theta$ between the director ${\bf n}$ of the
rod and the $z$ axis (Fig.~\ref{fig:model}), as 
\begin{equation}\label{eq2.1}
g(\theta) = \frac 1{MN_{\rm rod}} \sum_{j=1}^M\sum_{i=1}^{N{\rm
rod}} \delta\left(\theta_{i}(t_j)-\theta\right),
\end{equation}
where $M$ is the total number of time origins. A similar
function $g(\varphi)$ can be constructed for the azimuthal angle $\varphi$ in the
$xy$-plane. 

\begin{figure*}
\centering
\includegraphics[width=8.25cm]{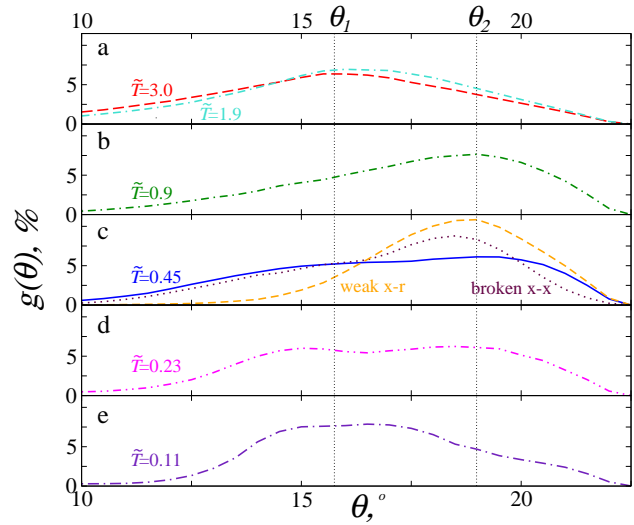}
\caption{Polar angle distribution $g(\theta)$ of rod orientation for the indicated reduced temperatures. The dotted vertical lines at $\theta_1\approx16^\circ$ and $\theta_2\approx19^\circ$ indicate the two preferable orientations of the rods. In panel c ($\w{T}=0.45$) the result for the system with weak x-r bonds (dashed line) and broken x-x bonds (dotted line) are shown as~well.}
\label{fig:theta}
\end{figure*}

\begin{figure*}
\centering
\includegraphics[width=8.25cm]{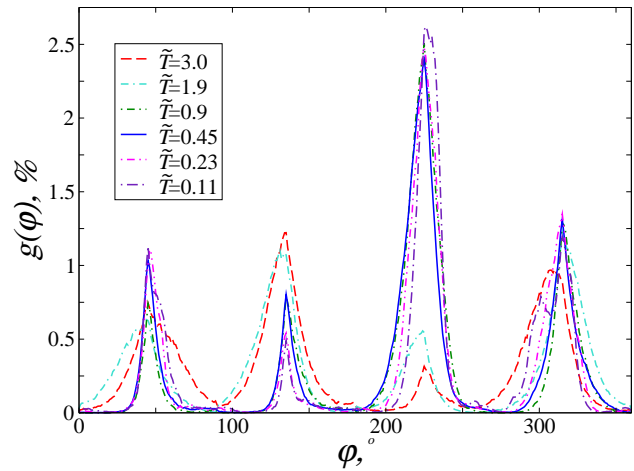}
\caption{Azimuthal angle distribution $g(\varphi)$ of rod orientation  for the indicated reduced temperatures. Notice the increase of the peak at $\varphi=225^\circ$ for $\w{T}\leq0.9$.}
\label{fig:phi}
\end{figure*}

The temperature dependence of the orientation distribution function $g(\theta)$  is shown in Fig.~\ref{fig:theta}. At high temperatures ($\w{T}=1.9,3.0$) a broad angular distribution of rods orientation peaked at $\theta_1\approx 16^\circ$ results from the sampling by the system large part of the available microstates. As temperature decreases, a second preferable orientation of the rods at $\theta_2\approx19^\circ$ appears, indicating some structural rearrangement below $\w{T}=1.9$ as seen at $\w{T}=0.9$. At lower temperatures both peaks become more pronounced, until at $\w{T}=0.11$ they merge into a ``plateau'' around $\theta_1$ and a peak at larger $\theta$ appears. A crossover between high and low temperature structures seems to occur between 1.9 and 0.9. This crossover is confirmed by looking at the azimuthal angle distribution $g(\varphi)$, shown in Fig.~\ref{fig:phi}. We see that while three peaks are present at all temperatures, the relatively sharp peak at $\varphi=225^\circ$ for $\w{T}\leq 0.9$ determines a prefered orientation of the rods in the $xy$-plane. Consequently, both distribution functions suggest the  appearance of additional order at low temperatures below $\w{T}=1.9$ as observed at $T\leq0.9$.

In order to single out the role of the xyloglucan network, we study two systems with broken x-x and with weak x-r bonds. The resulting distribution $g(\theta)$ for the system with broken x-x bonds is rather similar to the original one (Fig.~\ref{fig:theta}c).  However, for the system with weak x-r bonds, $g(\theta)$ has only one pronounced preferred orientation~$\theta_2$ (instead of being double-peaked), which suggests an easier reorientation of rods as for higher temperatures. 
The structure of the network of xyloglucan particles follows the structure of the rods as it is clearly illustrated in Fig.~\ref{fig:xyl_distr}, where one can recognize the seven peaks in the distribution of distances within a triplet for all studied temperatures (for example, we show $\w{T}=3.0;0.45$). For low temperatures the distribution is sharper, implying that  the network of xyloglucans is less flexible. From these results  we can conclude that the network of xyloglucan changes the optimal structure of the system, inducing glassy features like a broad, or double peaked, angular distribution.


\section{Slow relaxation}



\begin{figure*}
\centering
\includegraphics[width=8.25cm]{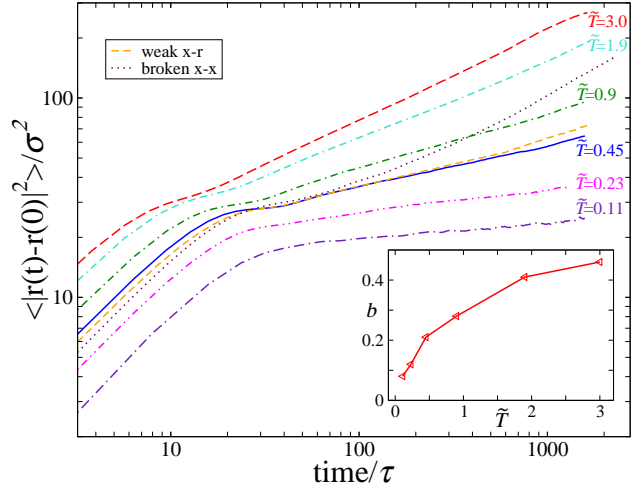}
\caption{MSD of rods in log-log scale. The initial part for all studied reduced temperatures is described by a power law with exponent $b\approx 0.8$. The long-time behaviour of the MSD is described by a power law with temperature dependent exponent $b(\w{T})$ given in the inset. Notice, that for all temperatures the dynamics is strongly subdiffusive ($b<1$). The two time regimes are separated by a ``plateau'' at $\w{T}=0.9;0.45$. The MSD for a system with broken x-x bonds is not well described by a power law for $t>30\tau$, while weakening of x-r bonds changes only slightly the MSD at $\w{T}=0.45$.}
\label{fig:msd}
\end{figure*}

We study the dynamics of our system by calculating the mean-square displacement (MSD) and the orientational time correlation function of the rods.
At all studied temperatures, we find that  the slow relaxation processes are well described by power laws and logarithmic functions. In Fig.~\ref{fig:msd} we show the time dependence of the ${\rm MSD} = \big\langle|{\bf r}(t)-{\bf r}(0)|^2 \big\rangle$ of rod beads in log-log scale. This dependence can be described by a power law
\begin{equation}\label{eq2.2}
\big\langle|{\bf r}(t)-{\bf r}(0)|^2 \big\rangle = a\cdot t^{b}
\end{equation}
with exponent $b \approx 0.8$ (instead of $b=2$ characteristic for the ballistic regime) for the initial part ($t<20\tau$) and with a temperature dependent $b(\w{T})$ for the slow regime ($t>30\tau$) shown in the inset of Fig.~\ref{fig:msd}. The diffusive regime ($b=1$) typical of liquids is never achieved in our simulations since, even at the highest temperature, the value of $b$ is less than 0.5, signaling a slow glassy dynamics of the rods. Moreover, the appearance of a ``plateau'' between the two regimes for $\w{T}=0.9$ corresponds to the emergence of another structurally arrested state, where the rods are trapped by their neighbours. This confirms that $\w{T}=0.9$ is below the crossover temperature between two glassy phases. By fitting $b(\w{T})$ (inset of Fig.~\ref{fig:msd}) to a power-law we find  $b(\w{T}) \propto \sqrt{\w{T}}$. Using this approximation, one can estimate the temperature typical of diffusive dynamics as $\w{T}\approx12$. We did not simulate such high temperature, where most xyloglucan bonds would be broken, since this regime is not interesting for cellulose. We found the following influence of xyloglucan network on the MSD at $\w{T}=0.45$:
weakening of x-r bonds changes only slightly the MSD, while  broken x-x interactions change 
qualitatively the MSD behaviour and seem to suppress the subdiffusive dynamics.



\begin{figure*}
\centering
\includegraphics[width=8.25cm]{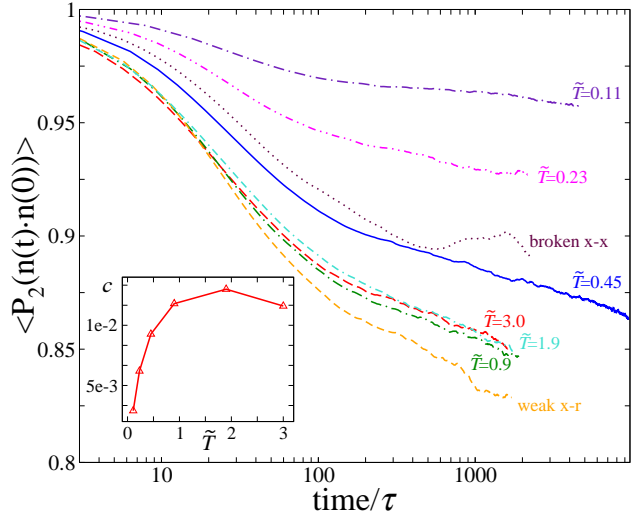}
\caption{Orientational time correlation function of the rod director $\bf n$ (Eq.~\ref{eq2.3}). The initial decay is exponential, whereas the long-time decay shows slow logarithmic relaxation. The breaking of x-x bonds (dotted line) at $\w{T}=0.45$ eliminates the slow dynamics, while the weakening of x-r interactions (short-dashed line) leads to a higher value of $c\approx0.017$ (Eq.~\ref{eq2.4}). Inset: the rate of change in rods orientation $c(\w{T})$.}
\label{fig:or}
\end{figure*}

Another quantity that can be used to characterize the glassy behaviour is the relaxation of rod orientation given  by  the following correlation function of the directors $\bf n$
\begin{equation}\label{eq2.3}
\big\langle P_2({\bf n}(t)\cdot{\bf n}(0)) \big\rangle = \frac
3{2MN_{\rm rod}} \sum_{j=1}^M\sum_{i=1}^{N_{\rm rod}}\big({\bf
n}_i(t_j)\cdot{\bf n}_i(t_j+t)\big)^2 - \frac12,
\end{equation}
where $P_2$ stands for the Legendre polynomial. From Fig.~\ref{fig:or} we notice that the above quantity has a behaviour similar to that of the 
MSD with two distinct relaxation processes: an initial fast exponential and slow logarithmic long-time decay. Thus, we fit these curves over the whole time interval with the following expression
\begin{equation}\label{eq2.4}
\big\langle P_2({\bf n}(t)\cdot{\bf n}(0) ) \big\rangle = A -
c\cdot\log t + p\cdot\exp(-t/\tau).
\end{equation}
The fitting parameter $c$ that characterizes the slow dynamics 
is shown as a function of temperature in the inset of Fig.~\ref{fig:or}. 
The curve  $c(\w{T})$ saturates for $\w{T}>0.9$, another sign of a structural rearrangement.  Above this temperature, we find an increase of broken x-x bonds from 2\% up to 6\%  at $\w{T}=3.0$.
The increased number of broken bonds at higher temperature makes the system 
less frustrated preventing further increase of $c$ with temperature. Nevertheless,
the slow logarithmic reorientation dynamics persists also above the crossover temperature.
Together with the data on strongly subdiffusive translational dynamics at all temperatures, 
this suggest a crossover between two types of structurally arrested glassy states like the one observed experimentally in a copolymer micellar system~\cite{chen:2003}.

The reorientation of rods depends also on the strength of the x-r 
interactions. Tenfold weakening of x-r interactions at $\w{T}=0.45$ results in a change of $\big\langle P_2({\bf n}(t)\cdot{\bf n}(0) ) \big\rangle$ more than that due to an increase of temperature to $\w{T}=3.0$ (Fig.~\ref{fig:or}) but slow logarithmic dynamics still exists. Conversely, breaking of x-x bonds ($D=0$)  leads to a non decaying amorphous behaviour of the  orientation of rods and eliminates the slow dynamics completely.


\section{Discussion and Conclusions}

We have studied by means of Molecular Dynamics simulations a model that captures the main structural features of cellulose. We have shown that, in a wide range of temperatures, a slow dynamics results from the competition between the microfibrils interactions and the stretchable network of xyloglucan molecules. 
The slow dynamics is characterized by logarithmic time dependence of orientational correlators and by
strongly subdiffusive dynamics of translational motion. 
These two features are robust and were observed for all temperatures investigated, 
around and below room temperature. Moreover, we have shown that weakening of the microfibril--xyloglucan interactions preserves the slow dynamics but influences the time scale of diffusion and reorentation.
These findings are compatible with the observation of logarithmic 
creep motion in plant cell walls, also when in presence of LTP that weakens 
the hydrogen bonds in the network~\cite{nieuwland:2005}. 
The loosening effect of LTP, rather than temperature, is thought to be one of the mechanisms that makes the extension of plant cell walls possible~\cite{nieuwland:2005}. Indeed also in our model, 
variations of temperature around room temperature do not affect noticeably the slow dynamics 
in the way a weakening of x-r interaction does. 

The complexity of our model produces an additional structural transition at low temperature. 
We find that a crossover beetween two glassy states, characterized by different angular distributions and parameters of slow dynamics. A similar glass-glass transition was found in copolymer micellar system~\cite{chen:2003}. We do not pursue further the study of the nature of this transition, that in our model occurs at low temperatures, not relevant for biological processes. However, the possibility of more than one structurally arrested state can be relevant for other related biopolymer networks.

\section*{Acknowledgements}

We are grateful to C. Mariani for critical reading of the manuscript. This work was sponsored by the Stichting Nationale Computerfaciliteiten (National Computing Facilities Foundation, NCF) for the use of supercomputer facilities, with financial support from the Netherlandse Organisatie voor Wetenschappelijk Onderzoek (Netherlands Organization for Scientific Research, NWO).

\clearpage


{\def\arraystretch{2}
\begin{table*}
\caption{Parameters of interaction potentials}
\centering
\begin{tabular}{lllll}\hline 
Bond & Potential & Energy & Length & Cut-off radius $R_{c,s}$ \\
\hline
b-b & $U_{\rm B}$, $U_{\rm LJ, b\text{-}b}$ & $A = 2500$, $\varepsilon_{\rm b\text{-}b}=10$ & $\sigma_{\rm b\text{-}b}=10$ &  $R_{c,\rm b\text{-}b}=15$  \\
r-r & $U_{\rm LJ, r\text{-}r}$ &  $\varepsilon_{\rm r\text{-}r}=0.001$ & $\sigma_{\rm r\text{-}r}=4$ & $R_{c,\rm r\text{-}r}=9$ \\
x-r & $U_{\rm LJ, x\text{-}r}$ &  $\varepsilon_{\rm x\text{-}r}=1\equiv \varepsilon$ & $\sigma_{\rm x\text{-}r}=1\equiv\sigma$ & $R_{c,\rm x\text{-}r}=2.3$   \\
x-x triplets & $U_{\rm M}$ & $D= 0.05$ & $r_{\rm eq}\in[20;60]$ &   $b=0.25\ \sigma^{-1}$ (Eq.~\ref{eq1.4}) \\
x-x all    & $U_{\rm LJ, x\text{-}x}$ & $\varepsilon_{\rm x\text{-}x}=0.1$ & $\sigma_{\rm x\text{-}x}=1$ &  $R_{c,\rm x\text{-}x}=1.12$  \\
\hline
\end{tabular}%
\label{tab1}
\end{table*}}

\end{document}